\documentclass{PoS}

\graphicspath{{plots/}} 
\usepackage{mystyle} 
\usepackage{amsmath}
\usepackage{color}
\let\normalcolor\relax
\newcommand{\ot}{\otimes}

\title{Non-perturbative renormalization for general improved staggered
bilinears}

\ShortTitle{NPR with Improved Staggered Bilinears}

\author{Andrew T. Lytle\\ School of Physics and Astronomy, University
of Southampton, Southampton SO17 1BJ, UK\\ E-mail:
        \email{a.t.lytle@soton.ac.uk}}

\author{\speaker{Stephen R. Sharpe}\\ 
        Physics Dept, University of Washington, Seattle WA 98195-1560\\
        E-mail: \email{srsharpe@uw.edu}}

\abstract{We present results for non-perturbative renormalization
(NPR) factors for staggered fermion bilinears of arbitrary spin and
taste. We use ``covariant'' bilinears which transform irreducibly
under the lattice translation and rotation group, and thus do not
mix. We form $\sim 30$ ratios which have no anomalous dimensions, and
compare the NPR results to those from
1-loop perturbation theory. We also compare
the absolute renormalization factors (which, in general,
do have anomalous dimensions) to 1-loop perturbation theory. We
use asqtad and HYP-smeared staggered valence fermions on the coarse
MILC asqtad lattices.}

\FullConference{The XXIV International Symposium on Lattice Field
			  Theory - LAT2011\\ July 10-17 2011\\ Lake
			  Tahoe, USA}

\begin{document}

\section{Introduction}
Precise knowledge of matching factors between lattice operators and
their continuum counterparts is necessary for phenomenological
applications of lattice QCD.  
In this work we present the status of our calculations of
matching factors using
non-perturbative renormalization (NPR)~\cite{NPR} as applied to valence
asqtad and HYP-smeared improved staggered fermions.
These actions have been used, respectively, in the calculation
of $m_s$ from light-hadron quantities~\cite{MILC1loop,MILC2loop,HPQCDms} 
and of $B_K$~\cite{BKPRD}.
In both of these calculations uncertainties in matching factors
are the dominant source of error.

Although NPR is widely used for other types of fermions, it has been
much less used for staggered fermions. We have chosen to
gain experience by first calculating matching factors for fermion
bilinears before proceeding to the four-fermion operators needed
for $B_K$. Specifically, we calculate matching factors for all
bilinears residing on a $2^4$ hypercube,
and make a detailed comparison with one-loop perturbation theory (PT).

The bilinears are also of interest in their own right
as part of a calculation of $m_s$. In particular,
with asqtad valence and sea quarks one finds
$m_s(\overline{\rm MS},2\;{\rm GeV}) = 76 \pm 8\;$MeV
using 1-loop matching~\cite{MILC1loop},
$87 \pm 6\;$MeV
with 2-loop matching~\cite{MILC2loop},\footnote{%
Here we are quoting older results based on the two lattice spacings
$a\approx 0.12\;$fm and $0.09\;$fm, since these are the spacings
we use in our calculation.
}
while replacing perturbative matching factors with
our preliminary NPR result gives
$103 \pm 3 ({\rm stat})$~\cite{Lytlethesis}.
One purpose of the present study is to determine the
systematic error in the latter result. This is important in order to
understand whether there is a significant deviation from the two-loop result.
We would also like to
check consistency with the more precise determination of $m_s$ using HISQ
fermions via the ratio $m_s/m_c$, which yields
$m_s(\overline{\rm MS},2\;{\rm GeV}) = 92.4\pm 1.5\;$MeV~\cite{HPQCDms}.

\section{Methodology}

We consider flavor non-singlet valence bilinears with
arbitrary Dirac matrix $\gamma_S$ and taste matrix
$\xi_F$. The matching or ``Z-factors'' appear in the relation
\begin{equation}
\overline{Q}(x)(\gamma_S\otimes\xi_F)Q'(x)
\cong
Z_{\gamma_{S} \otimes \xi_{F}}\sum_{A,B}
\overline{\chi}_A(n)\;
\overline{(\gamma_S\otimes\xi_F)}_{AB} 
U_{n+A,n+B}\; \chi_{B}'(n)
\,,
\end{equation}
where the symbol $\cong$ indicates that matrix elements
of the continuum operators on the left, evaluated with 
$\overline{\rm MS}$ regularization, equal those of
the lattice operators on the right, evaluated with lattice
regularization.
The continuum fields $Q$ have
four (implicit) tastes.
$\chi_B(n)$ is a staggered field
at a lattice position $n+B$ determined by the hypercube label, $n$,
and the offset within the hypercube, $B$.
$U_{n+A,n+B}$ is the gauge matrix connecting $\overline\chi$ to $\chi$,
and is constructed by averaging over the product of links along
all paths of minimal length.

We consider only bilinears with vanishing four-momentum,
which fall into 35 irreps under the lattice symmetry group
(as enumerated below).
While this proliferation of operators is often the bane of
staggered fermions, here it is a boon as there are 35 different
Z-factors to compare to PT.

One innovation we introduce is to use ``covariant'' bilinears.
Normally one sums over non-overlapping
positions of the hypercubes. This has the disadvantage that
the summed operators do not transform irreducibly under
lattice translations, and results in mixing between a subset
of the operators. Instead, if one sums over all translations (including
appropriate signs), one creates bilinears that live in irreps of the
full staggered translation and rotation group and thus can be shown
not to mix (at any order in PT).
A bonus is that it is simpler to implement such operators numerically.
More details will be given in Ref.~\cite{forthcoming}.

The different irreps are listed in Table~\ref{tab:bilins}.
They are organized according to the minimal number of gauge links
connecting $\overline\chi$ and $\chi$, and by their spin.
Matching factors of operators related by multiplication by
$(\gamma_5\otimes\xi_5)$ are the same, leading to the
relationships discussed in the caption.
In total there are 19 independent Z-factors for the 35 operators.

\begin{table}[hbt!]
\begin{center}
\begin{tabular}{c | l | l | l}
\hash{} links & S & V & T \\ \hline 
4 
&$\ct{\mathbf{1}}{\xi_{5}}$ 
&$\ct{\g_{\mu}}{\xi_{\mu} \xi_{5}}$ 
&\,\,$\ct{\g_{\mu} \g_{\nu}}{\xi_{\mu} \xi_{\nu} \xi_{5}}$ 
\\ 
3 
&$(\mathbf{1} \otimes \xi_{\mu} \xi_{5})$ &
 $(\g_{\mu} \ot \xi_{5}) \quad\, \ct{\g_{\mu}}{\xi_{\nu} \xi_{\rho}}$ &
 $\[\ct{\g_{\mu} \g_{\nu}}{\xi_{\mu} \xi_{5}} 
  \quad  \ct{\g_{\mu} \g_{\nu}}{\xi_{\rho}}\]$ 
\\ 
2
&$\ct{\mathbf{1}}{\xi_{\mu} \xi_{\nu}}$ 
&$(\g_{\mu} \ot \xi_{\nu}) \quad (\g_{\mu} \ot \xi_{\nu} \xi_{5})$ 
&$\[\textcolor{blue}{\ct{\g_{\mu} \g_{\nu}}{\mathbf{1}}}
  \quad \quad \,\,\, (\g_{\mu} \g_{\nu} \ot \xi_{5})\] 
  \quad (\g_{\mu} \g_{\nu} \ot \xi_{\nu} \xi_{\rho})$ 
\\ 
1
&$\ct{\mathbf{1}}{ \xi_{\mu}}$ 
&$\textcolor{blue}{\ct{\g_{\mu}}{\mathbf{1}}}
  \quad\,\,\,\, (\g_{\mu} \ot \xi_{\mu}\xi_{\nu})$ 
& $\[(\g_{\mu} \g_{\nu} \ot \xi_{\nu}) 
   \quad  (\g_{\mu} \g_{\nu} \ot \xi_{\rho} \xi_{5})\]$ 
\\
0
&$\textcolor{blue}{{\ct{\mathbf{1}}{\mathbf{1}}}}$
&$\ct{\g_{\mu}}{\xi_{\mu}}$
&\,\,$\ct{\g_{\mu} \g_{\nu}}{\xi_{\mu} \xi_{\nu}}$
\end{tabular}
\end{center}
\caption{Covariant bilinears forming irreps of the lattice symmetry
group. 
Indices $\mu$, $\nu$ and $\rho$ are summed from $1-4$, except
that all are different. Pseudoscalar and axial bilinears are not listed:
they can be obtained from scalar and vector, respectively, by
multiplication by $\gamma_5\otimes\xi_5$. Bilinears related
in this way have the same matching factors. This operation also implies
the identity of the Z-factors for the tensor bilinears
within square brackets.
Bilinears marked in \textcolor{blue}{blue} are 
used as the denominators of the ratios discussed in the text.}
\label{tab:bilins}
\end{table}

The only addition to the standard NPR methodology required for
staggered fermions is that one needs 16 lattice momenta for
each physical momenta, so that propagators and vertices are
all $16 \! \times \!16$ matrices. See Ref.~\cite{Lytlelat09} for more
details. We use momentum sources which allows us to work with
a small number of configurations (8-16). We calculate Z-factors
in the ${\rm RI}'$ scheme, and extrapolate linearly to
$am=0$. As is well known, this method fails for psuedoscalar
bilinears due to the pion pole. A more sophisticated
analysis is needed, and we do not present results
for these bilinears here.
We use $\sim 10$ different physical momenta for each
choice of quark masses. Our code is an adaptation of Chroma.

All our calculations use the MILC lattices generated with
the asqtad staggered fermion action and Symanzik glue.
Results presented here are from the ``coarse'' ensembles with
$a\approx 0.12\;$fm, size $20^3\times 64$, 
light sea-quark masses $am_\ell=0.01,0.02,0.03$ and strange sea-quark mass
$am_s=0.05$.
We are presently running on the $a\approx 0.09\;$fm ``fine''
ensembles and will present combined results in Ref.~\cite{forthcoming}.
For valence quarks we use both the asqtad action, with
unquenched masses $a m_{\rm val}=a m_{\ell}$,
and the HYP-smeared action, also with $a m_{\rm val}=a m_{\ell}$.
In the latter case the {\em physical} valence and sea quark masses differ,
since they are renormalized differently, but
this difference vanishes in the chiral limit.

One must also choose the links to be used in the bilinears.
We have tried various choices, but focus here on those giving
the best agreement with PT.
For asqtad valence quarks we use thick ``Fat~7 \!$+$\! Lepage'' links,
while for HYP valence quarks we use HYP-smeared links.

\section{Perturbative Predictions}

We compare our NPR matching factors with those from PT. The perturbative
results take the form:
\begin{equation}
Z_{\cal O}^{{\rm RI}',{\rm LAT}}(\mu,a)
=e^{-\int_{a(\mu_0)}^{a(\mu)} da \, \gamma_{\cal O}/\beta(a)} \,
Z_{\cal O}^{{\rm RI}',\overline{MS}}
\left(1 + a(\mu_0)[-2\gamma_{\cal O}^{(0)}\log(\mu_0 a)
+ C_{\cal O}^{\overline{\rm MS}}-C_{\cal O}^{\rm lat}]\right)
\,.
\label{eq:PTfull}
\end{equation}
Here $\mu$ is the scale of the NPR renormalization,
$a=\alpha/4\pi$, $\gamma_{\cal O}$ is the ${\rm RI}'$
anomalous dimension (known to 4 loops for S, P, V and A bilinears,
and to 3 loops for tensors), $\beta(a)$ is the corresponding
beta-function (known to four loops), and the
$C_{\cal O}$ are the finite parts of the one-loop matrix elements.
In words, the above equation says ``match (at 1-loop)
from lattice to $\overline{\rm MS}$
at scale $\mu_0$ (which we take to be our highest momentum
with $\mu_0\approx 3\;$GeV), then match (at 3-loop order)
from $\overline{\rm MS}$ to ${\rm RI}'$ in the continuum at scale $\mu_0$,
and finally run in the continuum down to the scale $\mu$.''

The one-loop lattice-$\overline{\rm MS}$ matching with asqtad and
HYP-smeared bilinears and the Symanzik gluon action has been worked
out in Ref.~\cite{KLS} for standard hypercube bilinears. 
We have extended this to covariant bilinears, finding that 
(a) the mixing between bilinears vanishes as it should and (b)
the diagonal mixing coefficients are the same as for the standard
bilinears. 
We have also done the calculation both
without and with mean-field improvement (MFI).

The PT predictions simplify considerably if we take ratios of
Z-factors having the same spin but different tastes, for then
the continuum matching and running cancels and the result
depends only on the lattice part of the matching calculation
\begin{equation}
\frac{Z_{\gamma_S\otimes\xi_{F1}}(p)}{Z_{\gamma_S\otimes\xi_{F2}}(p)}
=
1 + a(p)\left[C_{\gamma_S\otimes\xi_{F2}}^{\rm lat}-
C_{\gamma_S\otimes\xi_{F1}}^{\rm lat}\right] + {\cal O}(a^2)
\,.
\label{eq:PTratio}
\end{equation}
We take the denominators in our ratios to 
be the $Z-$factors of the taste singlet operators
(those shown in blue in Table~\ref{tab:bilins}).

\section{Results}

\begin{figure}
\begin{center}
\input{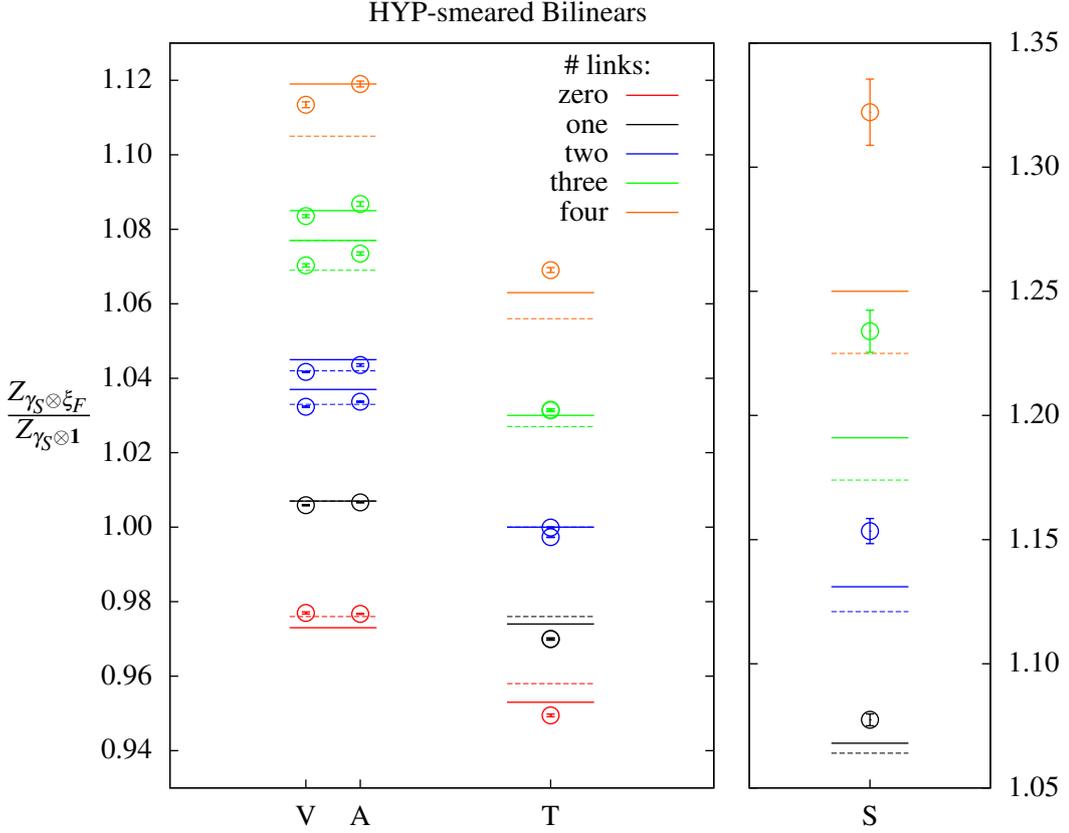}
\caption{Comparison of Z-factor ratios
for vector, axial, tensor and scalar bilinears to PT
for HYP fermions.  Horizontal lines show PT predictions, with
solid/dotted lines showing results with/without MFI.
Results are in the chiral limit for the momentum described in the
text. 
}
\label{fig:HYP_ratios}
\end{center}
\end{figure}

In Fig.~\ref{fig:HYP_ratios} we compare the ratios for V, A, T and S
bilinears composed of HYP-smeared quarks to PT.
Results are for momentum ``$(2,2,2,7)$'' in units of 
$(2\pi/L_s,2\pi/L_s,2\pi/L_s,2\pi/L_t)$, so that $|p|\approx 2\;$GeV.
We expect this momentum to be in the window, 
$\Lambda_{\rm QCD}\ll |p| \ll 1/a$,
where both non-perturbative corrections and lattice artifacts are small.
The color coding indicates the number of links in the operators in
the numerators of the ratios; the denominators have 1-link operators
for V and A, 2-link operators for T, and 0-link operators for S.

The level of agreement with 1-loop PT is quite different for V, A and T
bilinears than for scalars. For the former, PT is accurate
to $\sim 1\%$, capturing not only the
ordering with link number but also the ``fine structure'' within
a given link number. Note that statistical errors are very small,
despite the use of relatively few configurations.
Mean-field improvement leads to
slightly better agreement with PT, although the shifts are small.
An example of fine structure
is that the two 2-link tensor ratios
are predicted to be almost identical and are found to be very close.
Non-perturbative effects (which lead to the differences
between pairs of matched V and A bilinears) are at the subpercent level.
A striking example of this is that the two 1-link T ratios
are predicted to be the same and are indistinguishable on the
plot. The same holds for the two 3-link T ratios.

The agreement with PT is less good for the scalar bilinears.
While the correct ordering is obtained, the 1-loop predictions
differ by as much as $\sim 8\%$. This is, however, 
the difference that one would naively expect, 
because the missing two-loop term is of size $\alpha(p)^2\approx 0.08$.
We also note that we are less confident in the
linear chiral extrapolations for the S bilinears
than for the V, A and T bilinears.

\begin{figure}
\begin{center}
\input{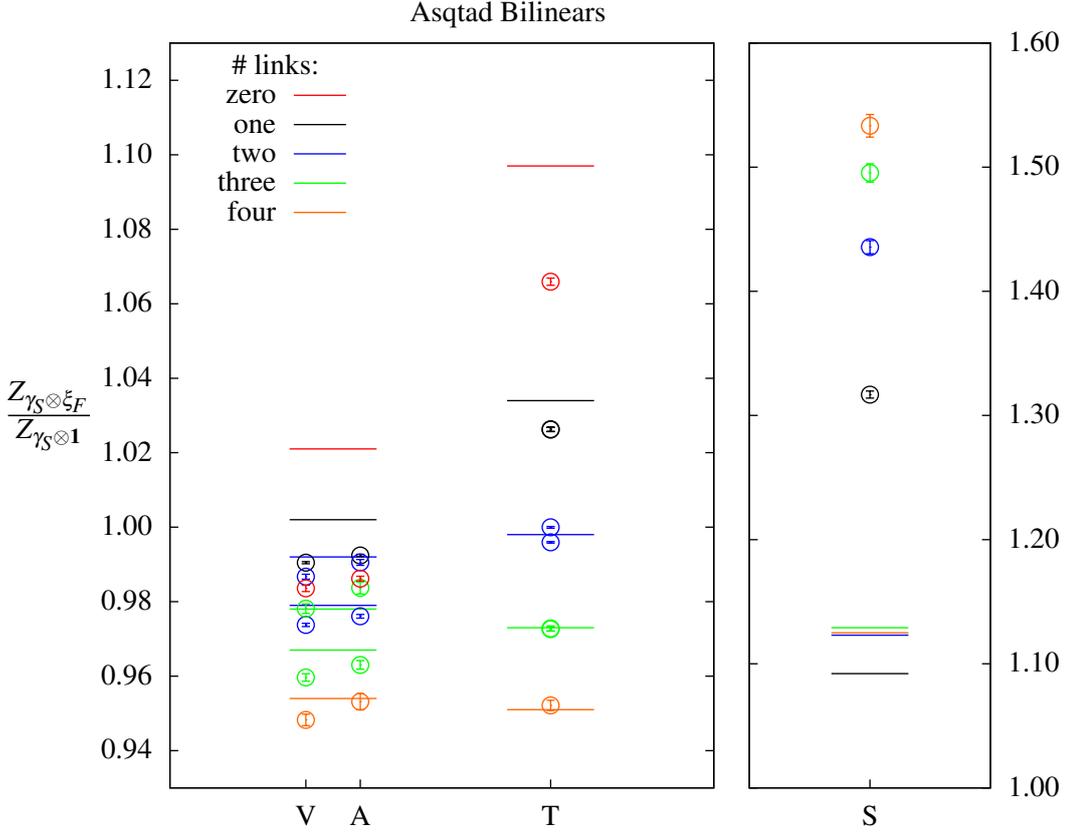}
\caption{Comparison of the vector, axial, tensor, and scalar ratios for asqtad fermions
to MFI PT.
Results are in the chiral limit for the momentum described in the
text. 
}
\label{fig:ASQ_ratios_MFI}
\end{center}
\end{figure}

Moving now to the asqtad bilinears, we find that mean-field
improvement is necessary to get reasonably accurate predictions.
This is not a surprise since, compared to HYP-smeared links,
the asqtad fat links have (normalized)
traces that are significantly further from unity.
We show the results in Fig.~\ref{fig:ASQ_ratios_MFI}.
For V, A and T bilinears,
the ordering with link number is predicted to be opposite
to that for HYP bilinears, and this is borne out in the NPR
results for tensors, and partially for the V and A bilinears.
Overall, MFI PT works at the $\sim 2\%$ level here.

By contrast, PT does very poorly for the asqtad scalar bilinears.
This is the reason why our NPR result for $m_s$ quoted above
lies significantly above those obtained using PT.
We are presently investigating the systematics associated with
the chiral and continuum extrapolations in order to firm up our
result.

\begin{figure}
\begin{center}
\includegraphics[width=0.7\textwidth]{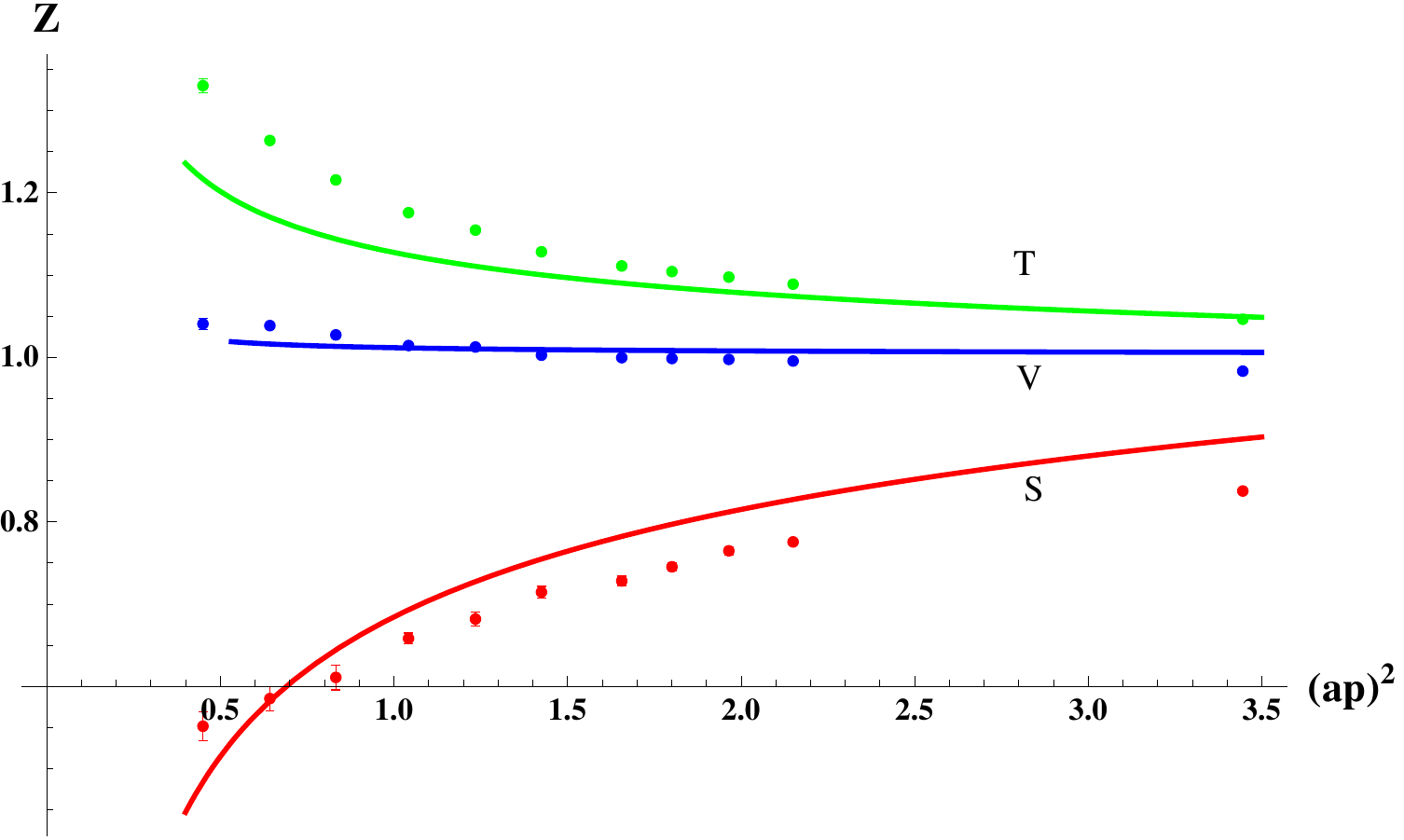}
\caption{Comparison of the scalar, vector, and tensor HYP denominators
for $m=0$ to perturbation theory.}
\label{fig:HYP_denoms}
\end{center}
\end{figure}

Finally we show in Fig.~\ref{fig:HYP_denoms} results for the three
denominators, i.e. $Z_{\gamma_S\otimes {\bf 1}}$ for spins S, V and T.
These are plotted vs. $(ap)^2$ and compared to the PT result
of Eq.~(\ref{eq:PTfull}). Note that all three operators
have non-vanishing anomalous dimensions in the ${\rm RI}'$
scheme, although that for the vector is very small.
We expect that non-perturbative
effects should be small for $p> 2\;$GeV, which translates to
$(ap)^2>1.4$. Indeed, we see that in this regime the NPR results
track the PT predictions fairly well, up to overall rescalings
of $\sim 5\%$.

\section{Conclusions and Outlook}

We have calculated all Z-factors of general ``covariant" bilinears
nonpertubatively and in one-loop PT for both HYP-smeared
and asqtad fermions.  
For HYP-smeared fermions, one-loop PT predictions for the ratios
give a good representation of the NPR results, with accuracy
varying from $\sim 1\%$ for V, A, and T bilinears to $\sim \alpha^2$
for scalars. For the absolute Z-factors the one-loop predictions
are also accurate to $\sim \alpha^2$ or better, and the predicted
running with $p$ is qualitatively described.
The latter result is encouraging for the calculation of $B_K$
with staggered fermions. This is because the tensor 
and scalar bilinears have anomalous dimensions which are
similar in magnitude to that of the operator needed for $B_K$.
If one-loop PT can represent these bilinears to an accuracy
of at least $\sim \alpha^2$ then it is reasonable to
expect the same to be true for the $B_K$ operators.
Thus the error estimate ($\delta B_K\approx \alpha(1/a)^2$) used
in the $B_K$ calculation is seen to be reasonable~\cite{BKPRD}.

We find PT to be less accurate for asqtad bilinears, even after
mean-field improvement, and to be very poor for
the scalar bilinear. The accuracy
gets considerably worse if one uses thin links in the operators.
This raises a concern that the result for $m_s$ obtained
using PT, albeit at 2-loop order, is suspect.
We hope to complete our calculation of $m_s$ using NPR on
two lattice spacings soon~\cite{forthcoming}.

\section*{Acknowledgements}
The work of AL is supported by STFC Grant ST/G000557/1.
The work of SS is supported in part by US DOE grant
no.~DE-FG02-96ER40956.
Computations were carried out on USQCD Collaboration 
clusters at Fermilab.
The USQCD Collaboration is
funded by the Office of Science of the U.S. Department of Energy.

\end{document}